\documentclass[manuscript]{aastex}

\usepackage{natbib}
\usepackage{color}

\slugcomment{Astrophysical Journal Letters}

\shorttitle{Macrospicule and Associated Jet by Kinked Flux Rope}
\shortauthors{Kayshap et al.}

\begin{document}


\title{Origin of Macrospicule and Jet in Polar Corona by A Small-scale 
Kinked Flux-Tube}


\author{P.~Kayshap\altaffilmark{1}}
\affil{Aryabhatta Research Institute of Observational Sciences (ARIES), Manora Peak, Nainital-263 129, India}

\author{Abhishek K. Srivastava\altaffilmark{1}}
\affil{Aryabhatta Research Institute of Observational Sciences (ARIES), Manora Peak, Nainital-263 129, India}

\author{K. Murawski\altaffilmark{2}}
\affil{Group of Astrophysics,UMCS, ul. Radziszewskiego 10, 20-031 Lublin, Poland}
\email{kmur@kft.umcs.lublin.pl}

\author{Durgesh Tripathi\altaffilmark{3}}
\affil{Inter-University Centre for Astrophysics, Post Bag - 4, Ganeshkhind, Pune-411007, India}
\email{durgesh@iucaa.ernet.in}

\begin{abstract}
We report an observation of a small scale flux-tube that undergoes kinking and triggers the macrospicule and a jet on November 11, 2010 in the north polar corona. The small-scale flux-tube emerged well before the triggering of macrospicule and as the time progresses the two opposite halves of this omega shaped flux-tube bent transversely and approached towards each other. After $\sim$ 2 minutes, the two approaching halves of the kinked flux-tube touch each-other and internal reconnection as well as energy release takes place at the adjoining location and a macrospicule was launched which goes upto a height of 12 Mm. Plasma starts moving horizontally as well as vertically upward along with the onset of macrospicule and thereafter converts into a large-scale jet which goes up to $\sim$ 40 Mm in the solar atmosphere with a projected speed of $\sim$ 95 km s$^{-1}$. We perform 2-D numerical simulation by considering the VAL-C initial atmospheric conditions to understand the physical scenario of the observed macrospicule and associated jet. The simulation results show that reconnection generated velocity pulse in the lower solar atmosphere steepens into slow shock and the cool plasma is driven behind it in form of macrospicule. The horizontal surface waves also appeared with the shock fronts at different heights, which most likely drove and spread the large-scale jet associated with the macrospicule.
\end{abstract}




\section{Introduction}

Macrospicules are giant spicules, mostly observed in polar coronal holes, 
reaching upto a height between 7{--}45~Mm above the solar-limb with a life-time 
of 3{--}45~minutes (e.g., Bohlin et al. 1975; Sterling 2000; Wilhelm 2000). A number 
of mechanisms have been proposed for the formation of such plasma ejecta, 
e.g., gas pressure pulse (Hollweg 1982), velocity pulse (Suematsu et al. 1982; Murawski et al. 2011). 
Shibata (1982) suggested that if reconnection takes place in upper 
chromosphere/lower corona (lower chromosphere/photosphere) macrospicule can be 
triggered due to magnetic reconnection (evolution of slow shocks). Alternative 
mechanisms have also been reported for the formation of such plasma ejecta
(e.g., Moore at al. 1977; Habbal \& Gonzalez 1991;
Mustsevoi \& Solovev 1997; De Pontieu et al. 2004; Kamio et al. 2010, and references cited
there)
 
Apart from the spicules and macrospicules in the lower solar atmosphere, 
the large-scale jets have significant role in mass and energy transport 
upto the higher corona, as well as in destabilizing large-scale coronal 
magnetic fields leading the eruptions (e.g., Innes et al. 1997;
Isobe \& Tripathi 2006; Chifor et al. 2006; Culhane et al. 2007; 2007; Filippov et al. 2009;
Tripathi et al. 2009; de Pontieu et al. 2010; De Pontieu et al. 2011; Judge et al. 2012).
Using STEREO/EUVI (Kaiser et al. 2008), a variety of 
solar jets on the basis of their sizes, life-times have been reported 
(Nistic\'o et al. 2009). The magnetic reconnection was found to be one of the 
drivers of coronal jets (Yokoyama et al. 1995; Innes et al. 1997;
Culhane et al. 2007; Chifor et al. 2008; Filippov et al. 2009; Nishizuka et al. 2009).
Pariat et al. (2009) have shown that reconnection 
generated nonlinear Alfv\'en waves can produce the polar coronal jets. 
Alternatively, the MHD pulse-driven models are also employed for the 
triggering of various large-scale solar jets (e.g.,
Srivastava \& Murawski 2011; Srivastava et al. 2012; Kayshap et al. 2013, and references
cited therein), and supported by 
recent observations (Morton et al. 2012). 

The inter-relationship of spicules and jets is also important to 
understand their formation processes. Kamio et al. (2010) have shown 
the association of jet with the macrospicule, which was triggered 
by a twisted magnetic flux-rope. Close association of the polar 
surges with macrospicules is also reported by Georgakilas et al. (2001)
. Moore et al. (2011) have suggested that granule-size emerging bi-poles 
(EBs) can trigger the longer spicules and associated Alfv\'en waves, 
while the larger EBs can form X-ray jets. In this letter, we report 
firstly the evidence of the activation of a small-scale bipolar 
twisted flux-tube in the lower polar corona, which undergoes internal 
reconnection and triggers a macrospicule and associated coronal jet. 
We study the relationship between the formation of macrospicule and 
associated jet, as well their most likely triggering 
mechanism using SDO/AIA observations and numerical simulation. 
In Section 2, we discuss the observations of macrospicule and associated jet. 
In Section 3, we present the results of numerical simulations. 
We outline discussion and conclusions in last section.

\section{Observations of the Macrospicule and Associated Jet}

High resolution observations (0.6~\arcsec per pixel with cadence 12 s) 
from the Atmospheric Imaging Assembly on board the Solar Dynamic Observatory (SDO)
, using its various filters sensitive to the plasma at different temperatures 
(AIA; Cheimets et al. 2009, DelZanna et al. 2011, Lemen et al. 2012, O'Dwyer et al. 2012),
recorded the origin of macrospicule and associated polar jet on 11 Nov 2010 
(cf., Figure 1 $\&$ MS-Jet-304.mpeg). 
For this study, we have used 304~{\AA} image sequence to study the triggering 
of macrospicule and jet that occurred at the north pole of the Sun. 


Figure~1 (top-panel) displays some selected snapshots of SDO/AIA 304~{\AA}  during the 
evolution of macrospicule and associated polar jet. A bipolar, omega-shaped, small-scale 
flux-tube is emerged at the polar cap around 00:58:08 UT (top-left panel). During the evolution 
of this flux-tube between 00:58:08{--}00:59:44~UT ($\sim$ 95 sec), the flux-tube shows some 
transverse bending of magnetic surfaces on its both the halves that may be the signature of 
the evolution of kink perturbations (cf., blue arrows in the snapshot 00:58:44 {--} 00:58:56 
UT, and cartoon in bottom panel). The combined effect of apparent rotation and 
transverse bending, generates the kinked flux-tube at smaller spatial-scales, which undergoes 
internal reconnection and further leads the macrospicule and jet (cf., MS-Jet-304.mpeg).
To the best of our knowledge, this is the first direct observation of the evolution of kink 
perturbations in the small-scale flux-tube, which further enables internal reconnection and 
leads the formation of macrospicule and associated jet that move along ambient open field lines 
(cf., Figure 2, and cartoon in Figure 1). The two halves and legs of the flux-tube come 
closer to each other and merge to produce a brightening (cf., snapshot on 00:59:44 UT),
which is very likely a signature of internal magnetic reconnection between the two opposite 
halves of a small-scale flux-tube. A macrospicule is triggered at the same time (cf., 01:00:56 UT 
snapshot in Figure 2, and cartoon in Figure 1). The macrospicule reaches up to a height 
of $\sim$12 Mm with a speed of about 80~km~s$^{-1}$, and faded within $\sim$ 4 minute. During 
this time, a jet like feature evolved which grew in vertical as well as horizontal direction
(cf., 01:00:56 UT {--} 01:02:08 UT snapshots in Figure 2). The detailed plasma dynamics can be 
seen in the associated movie MS-Jet-304.mpeg. We also investigated other high temperature 
filters of AIA (e.g. 171,211, 193 \AA\ ) where jet plasma fronts are clearly evident, however, 
the macrospicule material is only visible in 304 \AA\ channel (cf., MS-Jet-171.mpeg, 
MS-Jet-211.mpeg, MS-Jet-193.mpeg). Figure~3 displays a slit position along the macrospicule 
and jet (left panel) and corresponding height-time diagram (right panel). The two paths represented 
by pink and yellow colors respectively show the ejecting and down-flowing jet material. The 
denser core of the large-scale jet plasma goes up to a height $\sim$40~Mm with a speed of $\sim$ 
95 km~s$^{-1}$, while its some fainter traces are evident upto $\sim$ 50-60 Mm. In the first 
$\sim$10 minutes, the large-scale jet plasma material seen to be moving upward and later started 
to fall back towards the surface within total life-time of $\sim$ 24 minutes. The average outflow 
speed and acceleration were $\sim$ 95 km s$^{-1}$ and 89.015 m s$^{-2}$, while the average 
down-flow speed and downward acceleration are respectively  154 km s$^{-1}$ and -649.99 m s$^{-2}$.

Small-scale flux-tube in the polar corona, which is most-likely imposed by the kink 
perturbations due to asymmetric bending of magnetic surfaces on its both the halves, 
creates a reconnection diffusion region on the temporal scale of $\sim$ 95 sec. This 
region has length (2L) $\sim$ 12 Mm, that is the height of the emerged flux-tube at 
00:58:08 UT. The half-width (S) is $\sim$ 3.6 Mm that is the separation between the two 
opposite halves of the kinked flux-tube (cf., snapshot on 00:58:08 UT in Figure 1). 
Two opposite halves of the small-scale flux-tube approach to each-other with a 
supersonic speed of $\sim$ 76 km s$^{-1}$, (sound speed in choromosphere is 15 km 
s$^{-1}$) and reconnected. The height of the reconnection site is around $\sim$ 3.0-4.0 Mm 
in the polar region from the anchored footpoint of tube suggesting chromospheric reconnection which 
may trigger the evolution of slow shocks excited by the velocity pulse (Shibata 1982; Murawski et al. 2011; Kayshap et al. 2013)
and the observed plasma dynamics. Considering 
this as basic scenario, we numerically simulate the observed macrospicule and jet 
dynamics. 

\section{Numerical Model of the Macrospicule and Jet}

We perform a 2-D MHD numerical simulation to understand the physics of the 
observed macrospicule and jet using FLASH code (Lee \& Deane 2009) with an assumption that the polar corona 
is gravitationally stratified. The set of equations solved using FLASH code are as follows:

\begin{equation}
\label{eq:MHD_rho}
{{\partial \varrho}\over {\partial t}}+\nabla \cdot (\varrho{\bf V})=0\, ,
\end{equation}

\begin{equation}
\label{eq:MHD_V}
\varrho{{\partial {\bf V}}\over {\partial t}}+ \varrho\left ({\bf V}\cdot \nabla\right ){\bf V} =
-\nabla p+ \frac{1}{\mu}(\nabla\times{\bf B})\times{\bf B} +\varrho{\bf g}\, ,
\end{equation}

\begin{equation}
\label{eq:MHD_p}
{\partial p\over \partial t} + \nabla\cdot (p{\bf V}) = (1-\gamma)p \nabla \cdot {\bf V}\, ,
\end{equation}

\begin{equation}
\label{eq:MHD_B}
{{\partial {\bf B}}\over {\partial t}}= \nabla \times ({\bf V}\times{\bf B})\, , 
\hspace{3mm}
\nabla\cdot{\bf B} = 0\, .
\end{equation}

Here ${\varrho}$, ${V}$, ${B}$, $p = \frac{k_{\rm B}}{m} \varrho T$, $T$, $\gamma=5/3$,
${\bf g}=(0,-g)$ with its value $g=274$ m~s$^{-2}$, $m$, $k_{\rm B}$, are respectively 
the mass density, flow velocity, magnetic field, gas pressure, temperature,
adiabatic index, solar gravitational acceleration, mean particle mass, and Boltzmann's constant. 
Radiative cooling as well as thermal conduction are not included in our model as we are only interested
in the dynamics at this instance.
\subsection{Equilibrium Configuration}

We assume that the solar atmosphere is in the static equilibrium (${\bf V}_{\rm e}={\bf 0}$) 
with a force free magnetic field,
\begin{equation}\label{eq:B_e}
(\nabla\times{\bf B}_{\rm e})\times{\bf B}_{\rm e} = {\bf 0}\ , 
\end{equation}
\noindent
such that it satisfies the current free condition,
$\nabla\times{\bf B}_{\rm e}={\bf 0}$,
and it is specified by the magnetic flux function, $A$,
as
\begin{equation}\label{eq:B_e1}
{\bf B}_{\rm e}=\nabla \times (A\hat {\bf z})\, .
\end{equation}
Here the subscript e 
corresponds to equilibrium quantities.

We set a weakly curved arcade type magnetic field configuration by choosing
%
\begin{equation}
A(x,y) = B_{\rm 0}{\Lambda}_{\rm B}\cos{(x/{\Lambda}_{\rm B})} {\rm exp}[-(y-y_{\rm r})/{\Lambda}_{\rm B}]\, .
\end{equation}
%
%
%
%
Here, $B_{\rm 0}$ is the magnetic field at the reference level, which is initial location of the pulse $y=y_{\rm r}$, and the magnetic scale-height is
\begin{equation}
{\Lambda}_{\rm B}=2L/\pi\, .
\end{equation}
%
We set and hold fixed $L=200$ Mm and $y_{\rm r}$ = 10 Mm. 

As a result of Eq.~(\ref{eq:B_e}) 
the pressure gradient is balanced by the gravity force,
\begin{equation}
\label{eq:p}
-\nabla p_{\rm e} + \varrho_{\rm e} {\bf g} = {\bf 0}\, .
\end{equation}
%
%
%
With the ideal gas law and the $y$-component of 
Eq.~(\ref{eq:p}), we 
arrive at 
\begin{equation}
\label{eq:pres}
p_{\rm e}(y)=p_{\rm 0}~{\rm exp}\left[ -\int_{y_{\rm r}}^{y}\frac{dy^{'}}{\Lambda (y^{'})} \right]\, ,\hspace{3mm}
\label{eq:eq_rho}
\varrho_{\rm e} (y)=\frac{p_{\rm e}(y)}{g \Lambda(y)}\, ,
\end{equation}
where
\begin{equation}
\Lambda(y) = k_{\rm B} T_{\rm e}(y)/(mg)
\end{equation}
is the pressure scale-height, and $p_{\rm 0}$ denotes the gas 
pressure at the reference level that we choose in the solar corona at $y_{\rm r}=10$ Mm.

We take
an equilibrium temperature profile $T_{\rm e}(z)$ (cf., bottom-left panel of Figure 3)
for the solar atmosphere
that consists of VAL-C atmospheric model of Vernazza et al. (1981)
and obtain the corresponding gas pressure and mass density (not shown) using Eq.~(\ref{eq:pres}) .
In our simulation the transition region is located at $y\simeq 2.7$ Mm.
An extended corona and chromosphere is considered respectively above and below this 
with requirement of temperature minimum at $y\simeq 0.9$ Mm

\subsection{Perturbations}

We impulsively perturb the system in equilibrium by a Gaussian velocity pulse $V$
that is nearly parallel to the ambient magnetic field lines, viz.,
\begin{equation}
\hspace{-0.5cm}
V_{\parallel} (x,y,t=0) = A_{\rm v} 
\exp\left[ 
-\frac{(x-x_{\rm 0})^2 +(y-y_{\rm 0})^2} {w_{\rm}^2}
\right]
\, .
\label{eq:perturb}
\end{equation}
Here $A_{\rm v}$ is the amplitude of the pulse, $(x_{\rm 0},y_{\rm 0})$ is its initial position and
$w_{\rm }$ is its width. We set and hold fixed $A_{\rm v}=7.5$ 
km s$^{-1}$, 
$x_{\rm 0}=0$ Mm, $y_{\rm 0}= 0.9$ Mm, and $w=0.2$ Mm.

\subsection{Results of the Numerical Simulation}

We set the simulation box as 
(-10,10) Mm $\times$ (0,40) Mm and impose the boundary conditions by fixing
all plasma quantities to their equilibrium values in time for x- and y-directions,
while all plasma quantities remain invariant 
along the z-direction. In the study, we use AMR grid with minimum (maximum) level of 
refinement set to 3(8) (cf., Top-left panel of Figure~3). We launch the velocity pulse in the lower solar atmosphere, which is considered to be 
excited by the chromospheric reconnection generated energy between the opposite halves of the omega-shaped bipolar flux-tube (cf., Eq. 12).
As the pulse propagates in gravitationally stratified atmosphere, it converts into a slow shock at higher altitudes.
As a result, a low pressure region develops behind it and drives cool chromospheric plasma upwards. 
This lagging plasma exhibits the properties of the observed macrospicule.

Figure 5, displays key snapshots of the simulation results as the temperature-map 
(color) and velocity (arrows) of the plasma.  At 100 s (first image), the shock front of the initially 
launched pulse has reached upto $\sim$ 6~Mm. The chromospheric plasma, however, 
lags behind the shock front and has reached up to a height of $\sim$4~Mm. This lagging is 
primarily due to the rarefaction of the plasma behind the shock front. Rarefaction is the low pressure region
behind the shock front and due to the low pressure region surrounding plasma falls back in this region. The falling
back material pushed back due to the larger photospheric density, which is basically lagging of plasma.  
By the time the 
chromospheric plasma reaches up to a height of $\sim$10~Mm at 250~s (see the right-top 
panel), the shock front has reached a height more than 30~Mm. Our observational finding shows that 
macrospicule goes up-to $\sim$12~Mm with its  width $\sim$4~Mm and the life-time of $\sim$4~minutes. The simulation results presented 
here approximately mimic the observed dynamics of the macrospicule, which is the propagation of cool plasma.

At t=400 s, there are two bumps present on each side of the main macrospicule, which are 
very likely created by the secondary shocks. Due to these shocks the base of the spicule is 
getting wider, i.e., the surrounding material is moving upward along with the macrospicule, which is most likely the 
spreading plasma as observed in form of the initiation of jet. At this 
moment, the central part material is suppressed compared to the bumps due to the dominance 
of the down-flowing material at the center. The bumps on each side of the central part goes 
up-to $\sim$ 11 Mm while the central part suppressed up-to $\sim$ 8 Mm. 
The next snapshot (at t = 450 s), again shows that central region material is continuously falling back
towards the solar surface suppressing it while the side wise plasma are moving upward in the solar atmosphere and most likely form the core of jet. 
In observations, the macrospicule fades, and the large-scale jet plasma evolves around it, which may have striking similarity with this model result. This quasi-periodic rise and fall with minimum time-scale of 200 s of the cool chromospheric material is clearly visible over the longer duration in simulation
due to the arrival of wave trains of slow shocks (cf., Figure 4 and SIM-Jet.mpg). This time-scale though depends upon pulse-strength, steepening of slow shocks and their reflection from transition region creating down-flows. Moreover, the simulated plasma ejection creates horizontal surface waves at higher altitudes, which are clearly visible with comparatively high temperature shock fronts at different heights (cf., Figure 4 $\&$ Sim-Jet.mpg). These surface waves may interact with the various layers of the upper atmosphere
upto 40 Mm (cf., Fedun et al. 2011), and most likely  triggers the horizontal and vertical spread of the multi-temperature jet (cf., MS-Jet-171.mpeg, MS-Jet-211.mpeg, MS-Jet-193.mpeg). Quasi-periodic rise and fall of comparatively cooler (cf., MS-Jet-304.mpeg) macrospicule and jet may not be evident in the observational base-line as the upcoming pulse trains
may be observed by down-falling material (Srivastava \& Murawski 2011).

\section{Discussion and Conclusions}

Using the high resolution observations of SDO/AIA at 304~{\AA} , we studied 
the detailed evolution of a macrospicule and associated jet recorded on 11 November, 2010. In 
addition, we performed numerical simulation to qualitatively match with the observed macrospicule and jet using VAL-C 
model of the solar atmosphere and FLASH code. To the best of our knowledge, this is the first direct evidence of formation of a
macrospicule due to the magnetic reconnection between two opposite halves of 
an emerging small-scale, kinked bipolar loop at lower altitude in the polar corona. The
observed kinked small-scale flux-tube may also be a rotating helical structure that undergo in internal reconnection to trigger
the jet (e.g., Patsourakos et al. 2008; Nistic\'o et al. 2009; Pariat et al. 2009).
The triggering of the macrospicule was followed by the evolution of a large-scale jet. These types of kinked flux-rope, internal reconnection, and related plasma dynamics were only observed in the large-scale active regions, sometime leading to large-scale coronal eruptions, e.g. prominences, CMEs etc. (e.g., T{\"o}r{\"o}k \& Kliem 2004; Tripathi et al. 2007; Tripathi et al. 2009; Srivastava et al. 2010; Kliem et al. 2010; Srivastava et al. 2013a; Srivastava et al. 2013b, and references there in)
. However, the analogous conditions of the evolution of kinked bipolar-loop at small spatio-temporal scale in the polar corona is observed for the first time as episodic mechanism for the triggering of macrospicule and jet. The macrospicule goes up-to 
$\sim$ 12 Mm with the projected speed $\sim$ 80 km~s$^{-1}$, and had a life time of $\sim
$4~min. The brightened and core plasma of the large-scale jet goes up-to $\sim$ 40 Mm in the solar atmosphere with the
projected speed $\sim$ 95 km~s$^{-1}$, and its life time was $\sim$ 24 minute. The standard jet models deal with the direct reconnection
driven forces (j$\times$B) between open and closed fields lines in the corona leading to the jet plasma propulsion (Yokoyama et al. 1995; Nishizuka et al. 2009; Pariat et al. 2009). 
However, the present new episodic mechanism suggests that the internal reconnection in a small-scale loop in lower 
chromosphere further generates a velocity pulse that steepens in slow-shock wave trains propagating 
through ambient open field lines and triggering the dynamics of macrospicule and jet.   

Depending upon the height of reconnection site inside the chromosphere and amount of energy 
release during reconnection within small-scale flux-tube (e.g.,
Shibata 1982; Murawski et al. 2011; Kayshap et al. 2013, and references therein)
, it most likely generates the velocity pulse that 
further converts into a slow shock and exhibits the features of macrospicule and associated jet. The 
excitation of surface waves and the motion of the shock fronts and associated plasma may be 
responsible for the formation of the observed solar jet. Our numerical results, therefore, approximately 
and qualitatively match the observed plasma dynamics. We conclude that the kinking and chromospheric reconnection in 
the small-scale flux-tube can be an episodic mechanism to drive the observed macrospicule and associated jet via
secondary consequences in form of the evolution of velocity pulse and associated slow shocks.

\section{Acknowledgments}
We thanks reviewer for his/her valuable suggestions that improved the manuscript considerably.
We acknowledge the use of the SDO/AIA observations for this study. The data are provided 
courtesy of NASA/SDO, LMSAL, and the AIA, EVE, and HMI science teams. The FLASH code 
has been developed by the DOE-supported ASC/Alliance Center for Astrophysical 
Thermonuclear Flashes at the University of Chicago. AKS acknowledges Shobhna Srivastava 
for patient encouragement. KM thanks Kamil Murawski for his assistance in 
drawing numerical data. DT acknowledge the support from DST under Fast Track Scheme 
(SERB/F/3369/2012-2013), while AKS and PK acknowledge DST-RFBR Projet
(INT/RFBR/P-117).The computational resources were provided by the HPC
Infrastructure for Grand Challenges of Science and Engineering Project,
co-financed by the European Regional Development Fund under the
Innovative Economy Operational Program. PK acknowledges Jai Bhagwan for his encouragement and supports.


\clearpage

\begin{thebibliography}{99}
\bibitem[Bohlin et al. 1975] {Bohlin1975} Bohlin, J. D., Vogel, S. N., Purcell, J. D., Sheeley, N. R., Jr., Tousey, R., $\&$ Vanhoosier, M. E. 1975, ApJ, 197, L133
\bibitem[Cheimets et al. 2009] {Che2009} Cheimets, P., Caldwell, D. C., Chou, C., Gates, R., Lemen, J., Podgorski, W. A., Wolfson, C. J., $\&$ Wuelser, J.-P. 2009, in Society of Photo-Optical Instrumentation Engineers (SPIE) Conference Series, Vol. 7438, Society of Photo-Optical Instrumentation Engineers (SPIE) Conference Series
\bibitem[Chifor et al. 2006] {Chifor2006} Chifor, C., Mason, H. E., Tripathi, D., Isobe, H., $\&$ Asai, A. 2006, A$\&$A, 458, 965
\bibitem[2007] {Chifor2007} Chifor, C., Tripathi, D., Mason, H. E., $\&$ Dennis, B. R. 2007, A$\&$A, 472, 967
\bibitem[Chifor et al. 2008] {Chifor} Chifor, C., Young, P. R., Isobe, H., Mason, H. E., Tripathi, D., Hara, H., $\&$ Yokoyama, T. 2008, A$\&$A, 481, L57
\bibitem[Culhane et al. 2007] {Culhane2007} Culhane, L., et al. 2007, PASJ, 59, 751
\bibitem[De Pontieu et al. 2004] {DePon2004} De Pontieu, B., Erd\'eyi, R., $\&$ James, S. P. 2004, Nature, 430, 536
\bibitem[de Pontieu et al. 2010] {DePon2010} de Pontieu, B., et al. 2010, AGU Fall Meeting Abstracts, C3
\bibitem[De Pontieu et al. 2011] {DePon2011} De Pontieu, B., et al. 2011, Science, 331, 55
\bibitem[Del Zanna et al. 2011] {Zanna2011} Del Zanna, G., O'Dwyer, B., $\&$ Mason, H. E. 2011, A$\&$A, 535, A46
\bibitem[Fedun et al. 2011] {Fedun2011} Fedun, V., Shelyag, S., Erd\'elyi, R. 2011, ApJ, 727, 17
\bibitem[Filippov et al. 2009]{Filippov2009} Filippov, B., Golub, L., $\&$ Koutchmy, S. 2009, Sol. Phys., 254, 259
\bibitem[Georgakilas et al. 2001] {Georg2001} Georgakilas, A. A., Koutchmy, S., $\&$ Christopoulou, E. B. 2001, A$\&$A, 370, 273
\bibitem[Habbal $\&$ Gonzalez 1991] {Habbal1991} Habbal, S. R., $\&$ Gonzalez, R. D. 1991, ApJ, 376, L25
\bibitem[Hollweg 1982] {Holl1982} Hollweg, J. V. 1982, ApJ, 257, 345
\bibitem[Innes et al. 1997] {Innes1997} Innes, D. E., Inhester, B., Axford, W. I., $\&$ Wilhelm, K. 1997, Nature, 386, 811
\bibitem[Isobe $\&$ Tripathi 2006] {Isobe2006} Isobe, H., $\&$ Tripathi, D. 2006, A$\&$A, 449, L17
\bibitem[Judge et al. 2012] {Judge2012} Judge, P. G., de Pontieu, B., McIntosh, S. W., $\&$ Olluri, K. 2012, ApJ, 746, 158
\bibitem[Kaiser et al. 2008] {Kaiser2008} Kaiser, M. L., Kucera, T. A., Davila, J. M., St. Cyr, O. C., Guhathakurta, M., $\&$ Christian, E. 2008, Space Sci. Rev., 136, 5
\bibitem[Kamio et al. 2010] {Kamio2010} Kamio, S., Curdt, W., Teriaca, L., Inhester, B., $\&$ Solanki, S. K. 2010, A$\&$A, 510, L1
\bibitem[Kayshap et al. 2013] {Kayshap2012} Kayshap, P., Srivastava, A. K., $\&$ Murawski, K. 2013, ApJ, 763, 24
\bibitem[Kliem et al. 2010] {Kliem2010} Kliem, B., Linton, M. G., T{\"o}r{\"o}k, T., $\&$ Karlick{\'y} M. 2010, Sol. Phys., 266, 91
\bibitem[Lee $\&$ Deane 2009] {Lee2009} Lee, D., $\&$ Deane, A. E. 2009, Journal of Computational Physics, 228, 952
\bibitem[Lemen et al. 2012] {Lem12} Lemen, J. R., et al. 2012, Sol. Phys., 275, 17
\bibitem[Moore et al. 2011] {Moore2011} Moore, R. L., Sterling, A. C., Cirtain, J. W., $\&$ Falconer, D. A. 2011, ApJ, 731, L18%
\bibitem[Moore at al. 1977] {Moore1977} Moore, R. L., Tang, F., Bohlin, J. D., $\&$ Golub, L. 1977, ApJ, 218, 286
\bibitem[Morton et al. 2012] {Mor2012} Morton, R. J., Srivastava, A. K., $\&$ Erd\'eyi, R. 2012, A$\&$A, 542, A70
\bibitem[Murawski et al. 2011] {Mura2011} Murawski, K., Srivastava, A. K., $\&$ Zaqarashvili, T. V. 2011, A$\&$A, 535, A58
\bibitem[Mustsevoi $\&$ Solovev 1997] {Must1997} Mustsevoi, V. V., $\&$ Solov’ev, A. A. 1997, Astronomy Reports, 41, 219
\bibitem[Nishizuka et al. 2009] {Nisi2009} Nishizuka, N., $\&$ Shibata, K. 2009, in Astronomical Society of the Pacific Conference
Series, Vol. 415, The Second Hinode Science Meeting: Beyond Discovery-Toward
Understanding, ed. B. Lites, M. Cheung, T. Magara, J. Mariska, $\&$ K. Reeves, 188
\bibitem[Nistic\'o et al. 2009] {Nistic2009} Nistic\'o, G., Bothmer, V., Patsourakos, S., $\&$ Zimbardo, G. 2009, Sol. Phys., 259, 87
\bibitem[O'Dwyer et al. 2010] {ODwyer2010} O'Dwyer, B., Del Zanna, G., Mason, H. E., Weber, M. A., $\&$ Tripathi, D. 2010, A$\&$A, 521, A21
\bibitem[Pariat et al. 2009] {Par2009} Pariat, E., Antiochos, S. K., $\&$ DeVore, C. R. 2009, ApJ, 691, 61
\bibitem[Patsourakos et al. 2008] {pat2008} Patsourakos, S., Pariat, E., Vourlidas, A., $\&$ Antiochos, S.K., 2008, ApJ, 680,73
\bibitem[Shibata 1982] {Shib1982} Shibata, K. 1982, Sol. Phys., 81, 9
\bibitem[Srivastava et al. 2013a] {Abhi2013} Srivastava, A. K., Botha, G. J. J., Arber, T. D., $\&$ Kayshap, P. 2013a, ArXiv e-prints (2013arXiv1301.2927S)
\bibitem[Srivastava et al. 2012] {Abhi2012} Srivastava, A. K., Erd\'eyi, R., Murawski, K., $\&$ Kumar, P. 2012, Sol. Phys., 166
\bibitem[Srivastava et al. 2013b]{Sri13b} Srivastava, A. K., Erd\'eyi, R., Tripathi, D., Fedun, V., Joshi, N. C., $\&$ Kayshap, P. 2013b,
ApJ, 765, L42
\bibitem[Srivastava $\&$ Murawski 2011] {Abhis2011} Srivastava, A. K., $\&$ Murawski, K. 2011, A$\&$A, 534, A62
\bibitem[Srivastava et al. 2010] {Abhi2010} Srivastava, A. K., Zaqarashvili, T. V., Kumar, P., $\&$ Khodachenko, M. L. 2010, ApJ, 715, 292
\bibitem[Sterling 2000] {Sterling2000} Sterling, A. C. 2000, Sol. Phys., 196, 79
\bibitem[Suematsu et al. 1982] {Sumo1982} Suematsu, Y., Shibata, K., Neshikawa, T., $\&$ Kitai, R. 1982, Sol. Phys., 75, 99
\bibitem[T{\"o}r{\"o}k $\&$ Kliem 2004] {Torok2004} T{\"o}r{\"o}k, T., $\&$ Kliem, B. 2004, in ESA Special Publication, Vol. 575, SOHO 15 Coronal Heating, ed. R. W. Walsh, J. Ireland, D. Danesy, $\&$ B. Fleck, 56
\bibitem[Tripathi et al. 2009] {DT2009} Tripathi, D., Gibson, S. E., Qiu, J., Fletcher, L., Liu, R., Gilbert, H., $\&$ Mason, H. E. 2009, A$\&$A, 498, 295
\bibitem[Tripathi et al. 2007] {DT2007} Tripathi, D., Solanki, S.K., Mason, H.E., $\&$ Webb, D.F. 2007, A$\&$A, 472, 633
\bibitem[Vernazza et al. 1981] {Ver81}  Vernazza, J. E., Avrett, E. H., $\&$ Loeser, R. 1981, ApJS, 45, 635
\bibitem[Wilhelm 2000] {Wil2000} Wilhelm, K. 2000, A$\&$A, 360, 351
\bibitem[Wuelser et al. 2004] {Wuelser} Wuelser, J.-P., et al. 2004, in Society of Photo-Optical Instrumentation Engineers (SPIE) Conference Series, Vol. 5171, Society of Photo-Optical Instrumentation Engineers (SPIE) Conference Series, ed. S. Fineschi $\&$ M. A. Gummin, 111
\bibitem[Yokoyama et al. 1995] {Yoko1995} Yokoyama, T., Shibata, K. 1995, Nature, 375, 42
\end{thebibliography}

\clearpage
\begin{figure*}
\vspace{-0.8cm}
\hspace{2.3cm}
\includegraphics[scale=0.30]{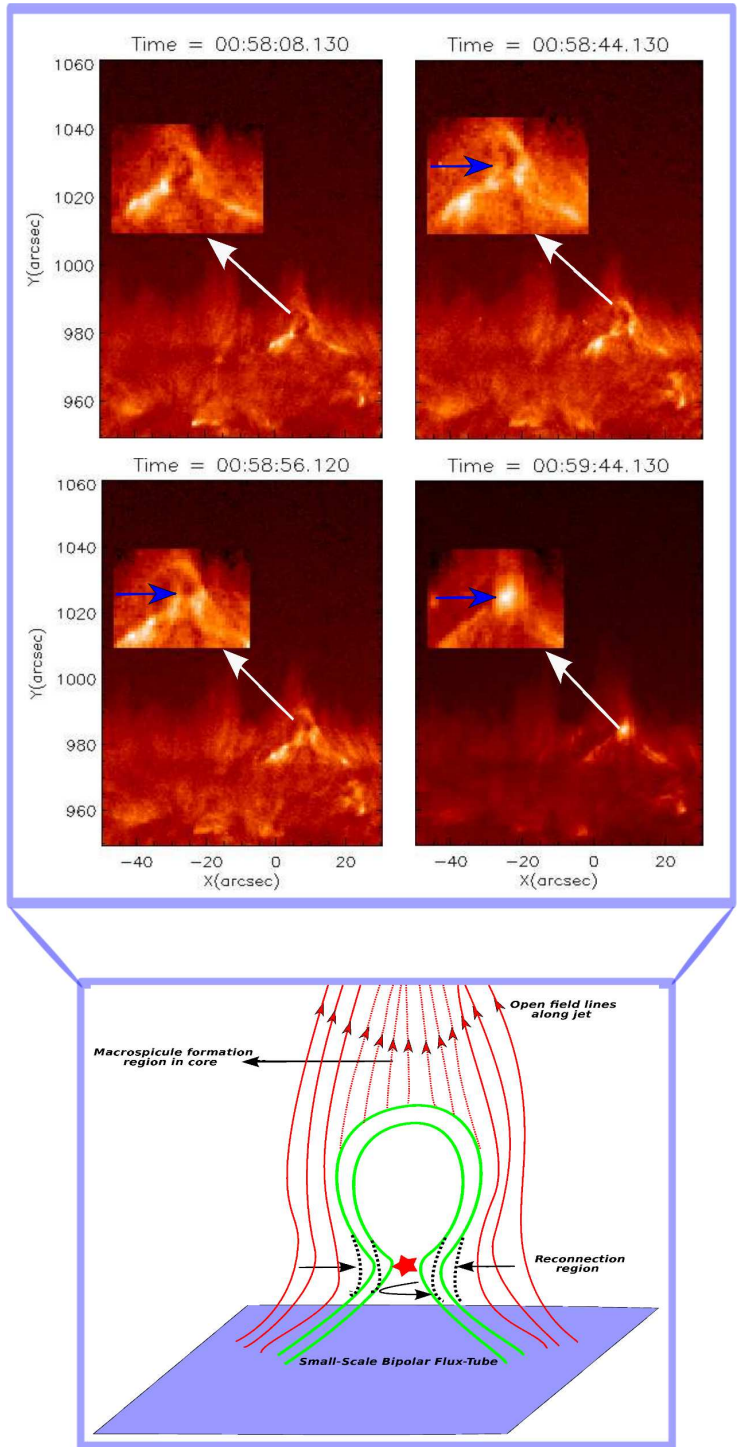}
\vspace{-0.8cm}
\caption{Top-panel: SDO/AIA 304 \AA\ ($\sim$ 10$^{5}$ K) snapshots show the evolution of a small-scale flux-tube in the polar region to trigger the macrospicule and associated jet. Enlarged view of the small-scale, omega shaped bipolar loop shows that the two opposite halves of the kinked flux-tube approach each other and reconnect internally (00:59:44 UT) to release the energy in the chromosphere. Bottom panel: A schematic showing the formation of transverse bending and apparent rotation in small-scale flux-tube that cause its kinking to trigger the internal reconnection and energy release.
}
\label{fig:1}
\end{figure*}
\clearpage
\begin{figure*}
\mbox{
\includegraphics[scale=0.7]{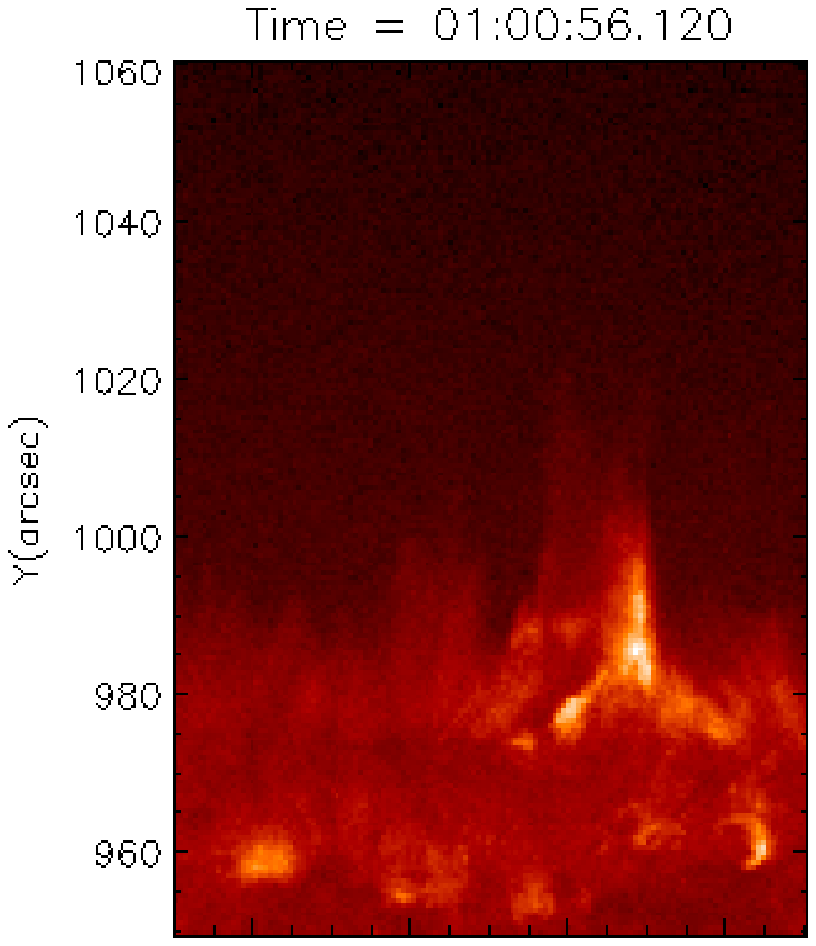}
$ \color{white} \put(-60,60){\vector(1,0){20}}\put(-120,63){Macrospicule}$
$ \color{white} \put(-88,120){\vector(1,-1){25}}\put(-110,125){Plasma Motion}$
\hspace{-0.4cm}
\includegraphics[scale=0.7]{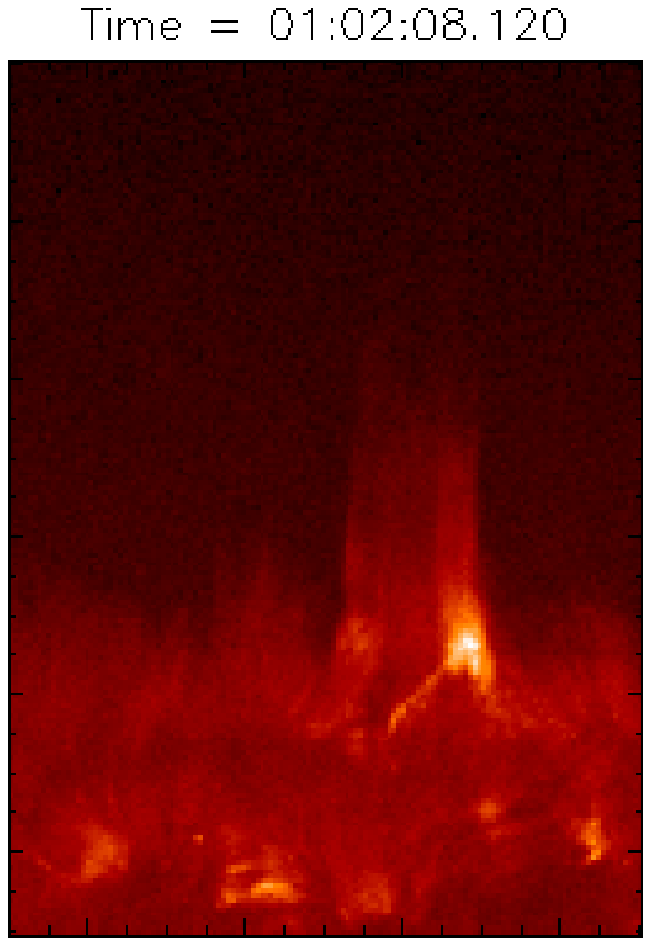}
$ \color{white} \put(-67,60){\vector(1,0){20}}\put(-130,65){Spicule Fading}$
$ \color{white} \put(-80,130){\vector(1,-1){25}}\put(-110,135){Jet Plasma}$
\hspace{-0.4cm}
\includegraphics[scale=0.7]{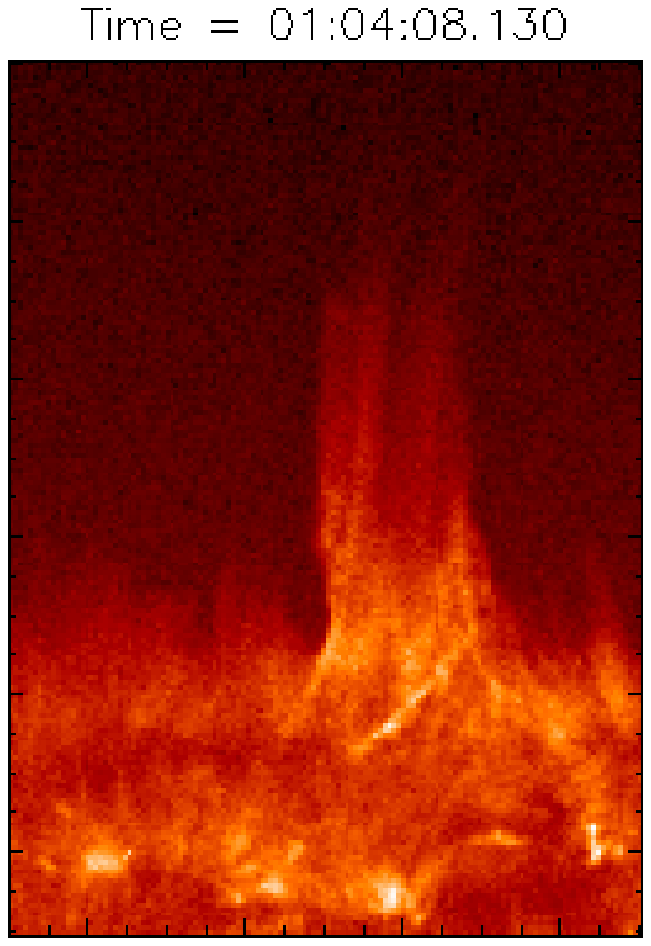}
}
\mbox{
\includegraphics[scale=0.7]{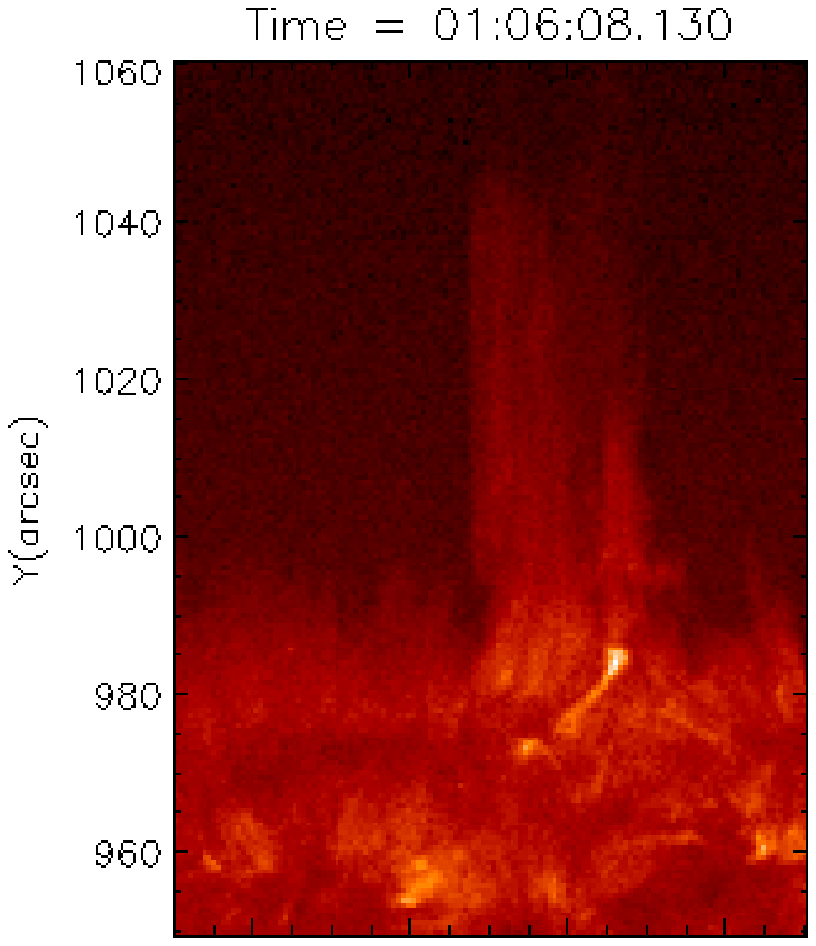}
\includegraphics[scale=0.7]{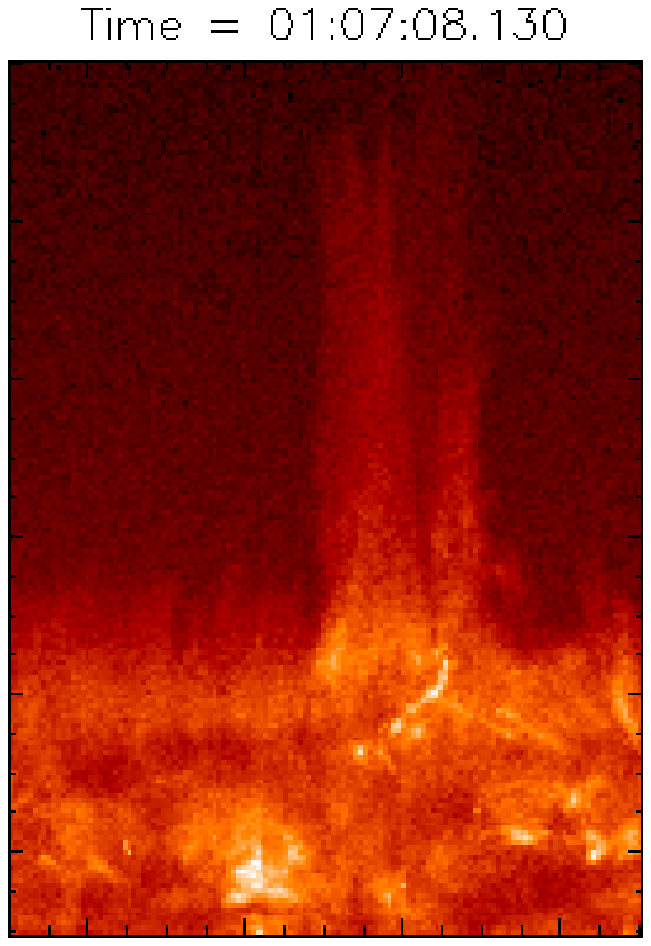}
\includegraphics[scale=0.7]{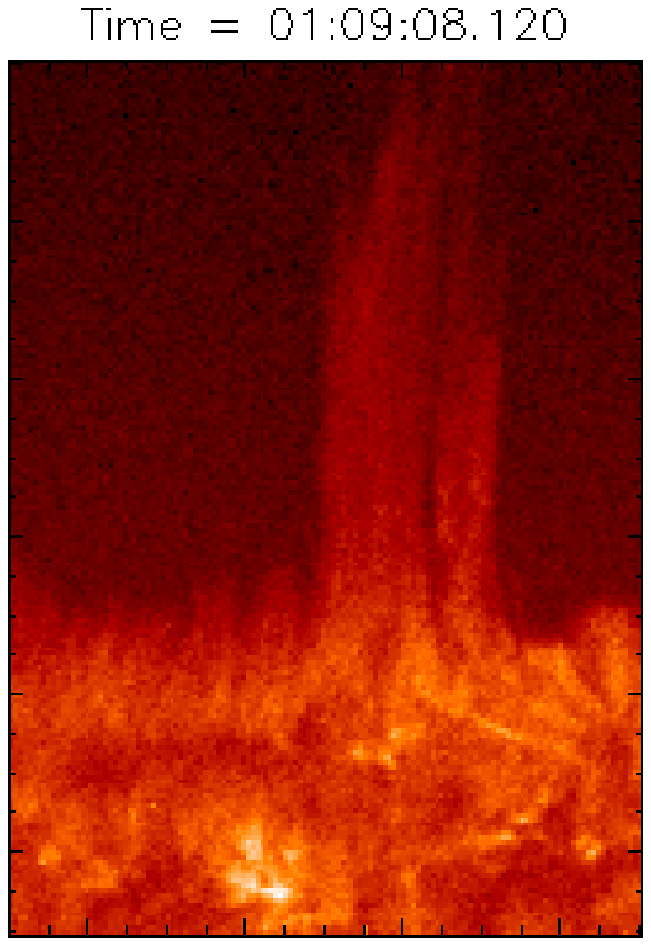}
}
\mbox{
\includegraphics[scale=0.7]{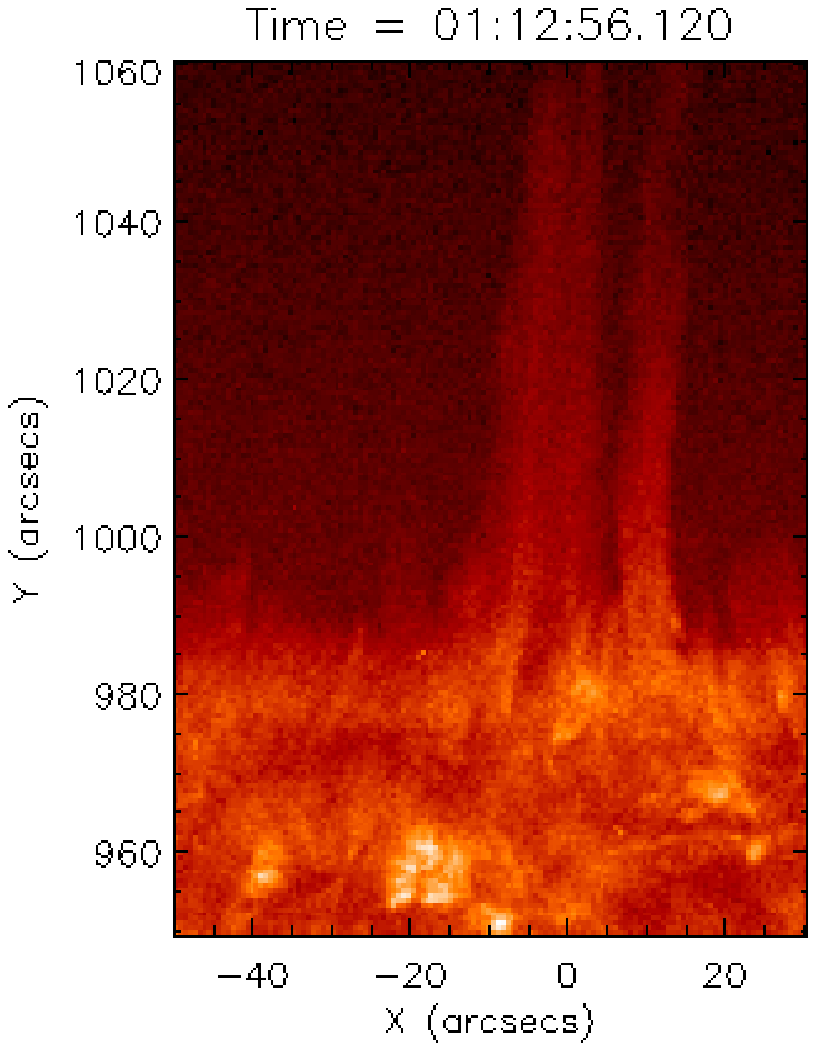}
\includegraphics[scale=0.7]{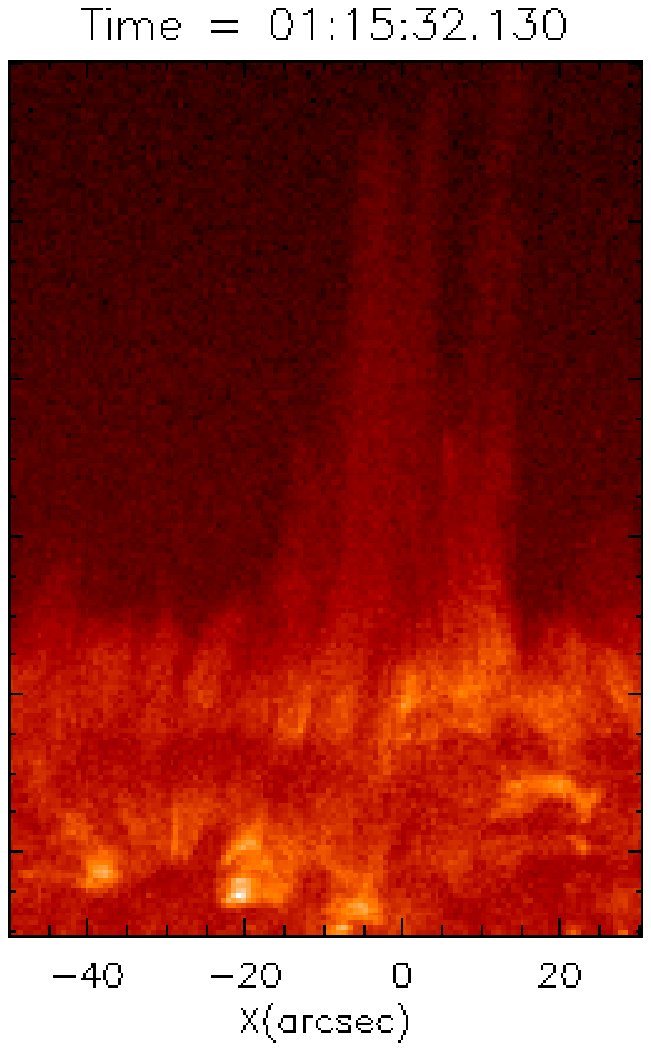}
$\color{white} \put(-95,170){\vector(1,-1){25}}\put(-130,175){Down-falling Plasma}$
\hspace{-0.25cm}
\includegraphics[scale=0.7]{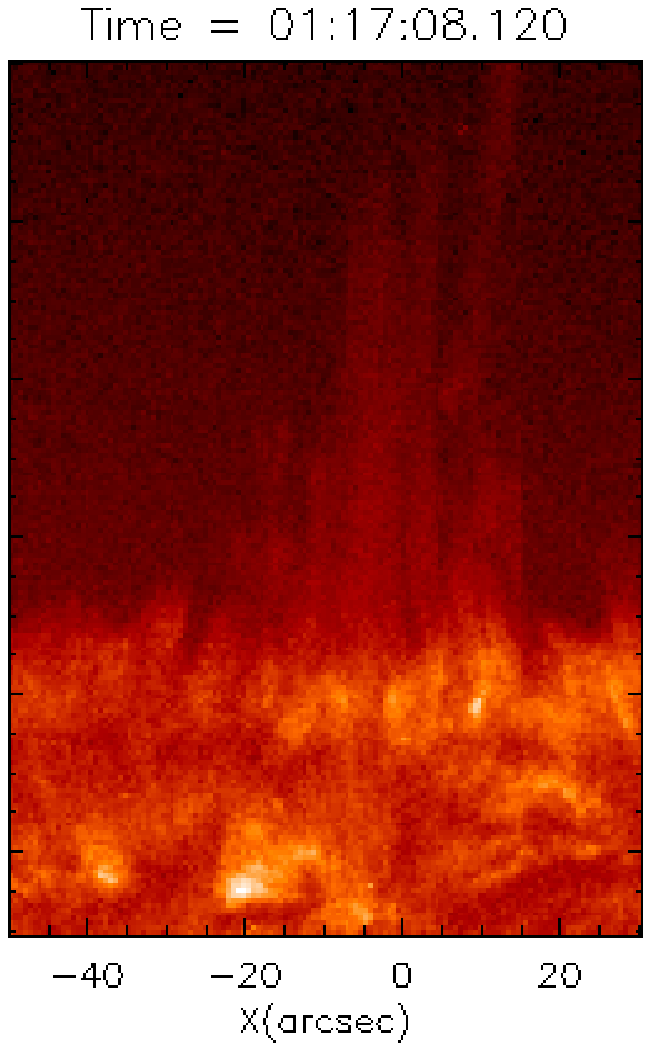}
}
\vspace{-0.8cm}
\caption{\small The macrospicule triggers on 00:59:44 UT due to the reconnection between opposite halves of flux-tube as shown in Figure 1, and finally converts into a jet (cf., 01:02:08-01:06:08 UT snapshot). The comprehensive dynamics of the macrospicule and jet are shown in the attached movie MS-Jet-304.mpeg. The cartoon in Figure 1 also depicts the scenario of the formation of macrospicule and jet along ambient open field lines.}
\label{fig:1}
\end{figure*}
\clearpage
\begin{figure*}
\mbox{
\hspace{-6.0cm}
\includegraphics[scale=0.8]{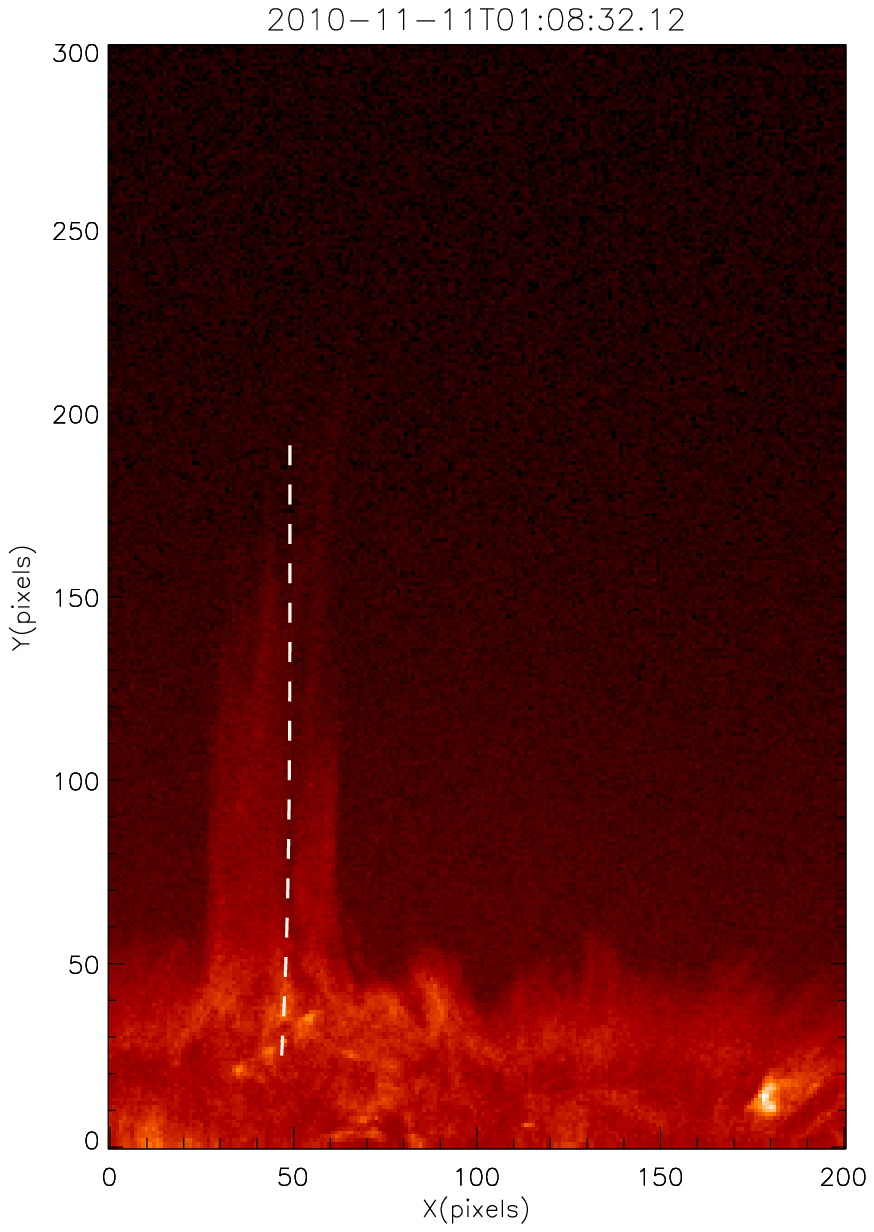}
\hspace{-4.0cm}
\includegraphics[scale=0.8]{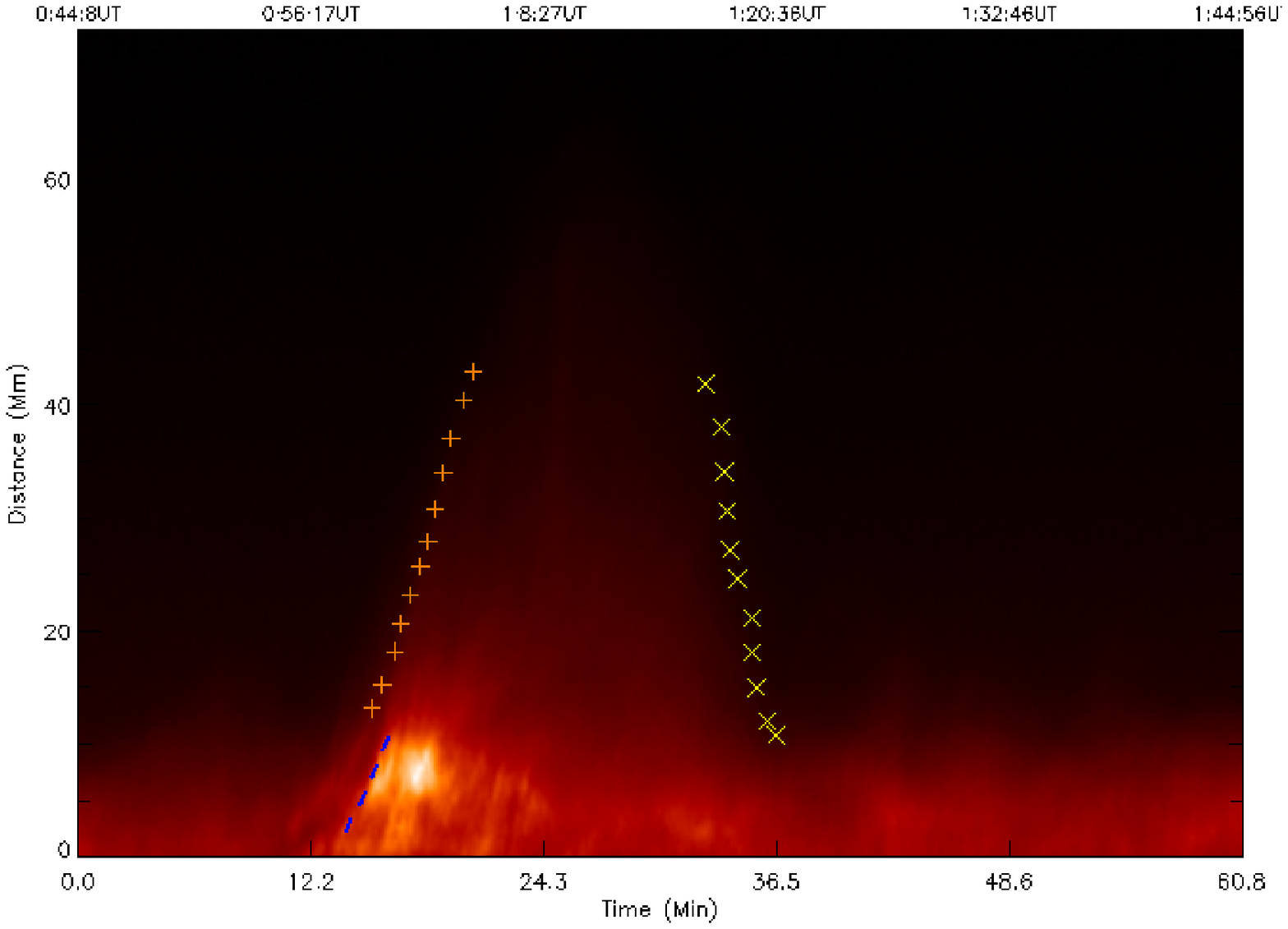}
}
\caption{Left image shows the location of the slit along the jet to measure the height-time profile. Right snapshot shows the height-time profile of the jet. Two different paths on this image show respectively the acceleration of the  jet (red path) as well as its deceleration (yellow path). The bright and denser part of the jet goes up-to $\sim$ 40 Mm with projected speed of $\sim$ 95 km s$^{-1}$, while some of its fainter traces are also observed upto 50-60 Mm.
}
\label{fig:2}
\end{figure*}
\clearpage
\begin{figure*}
\begin{center}
\mbox{
\vspace{-10.0cm}
\includegraphics[scale=0.5]{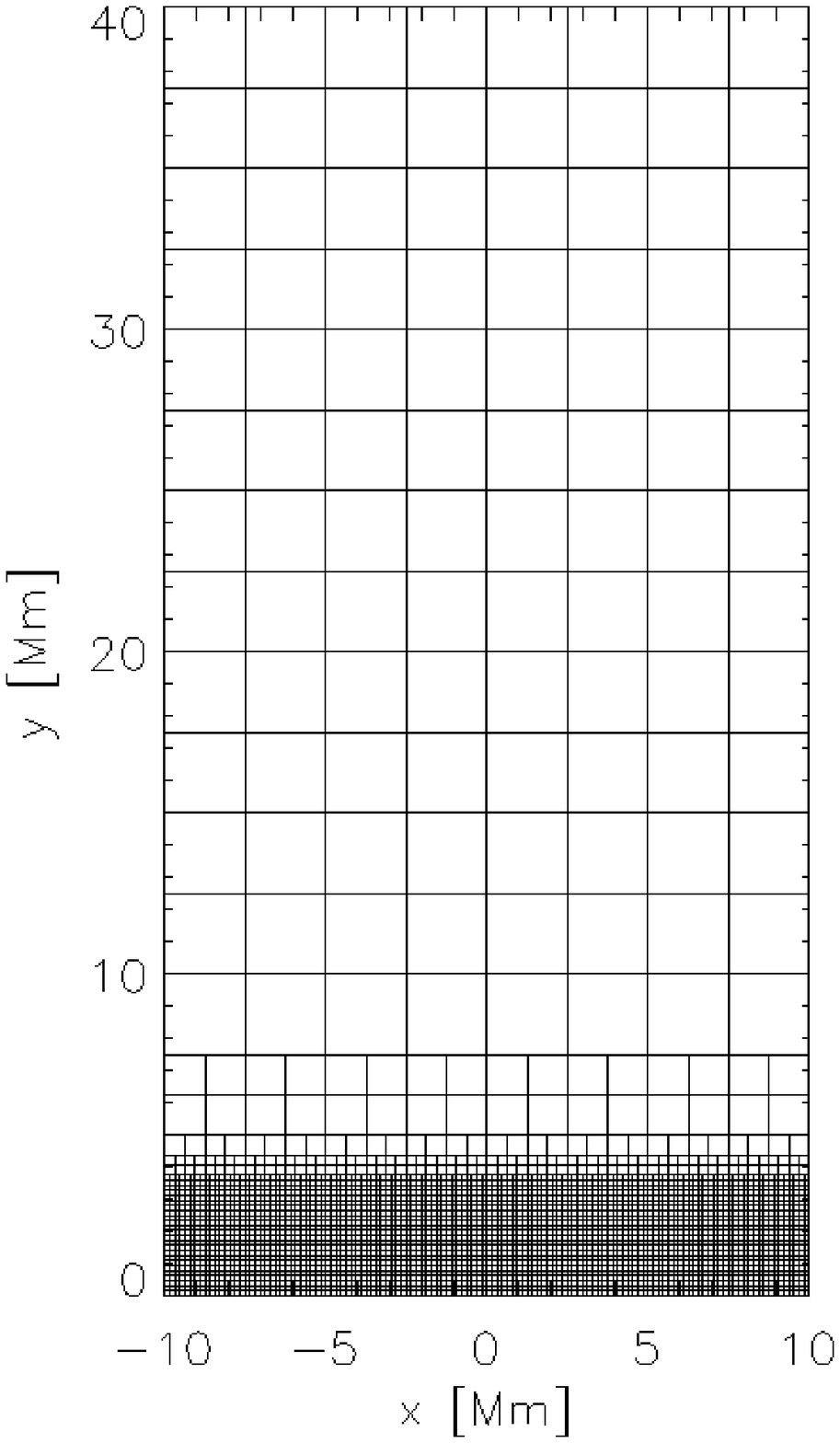}
\includegraphics[scale=0.5]{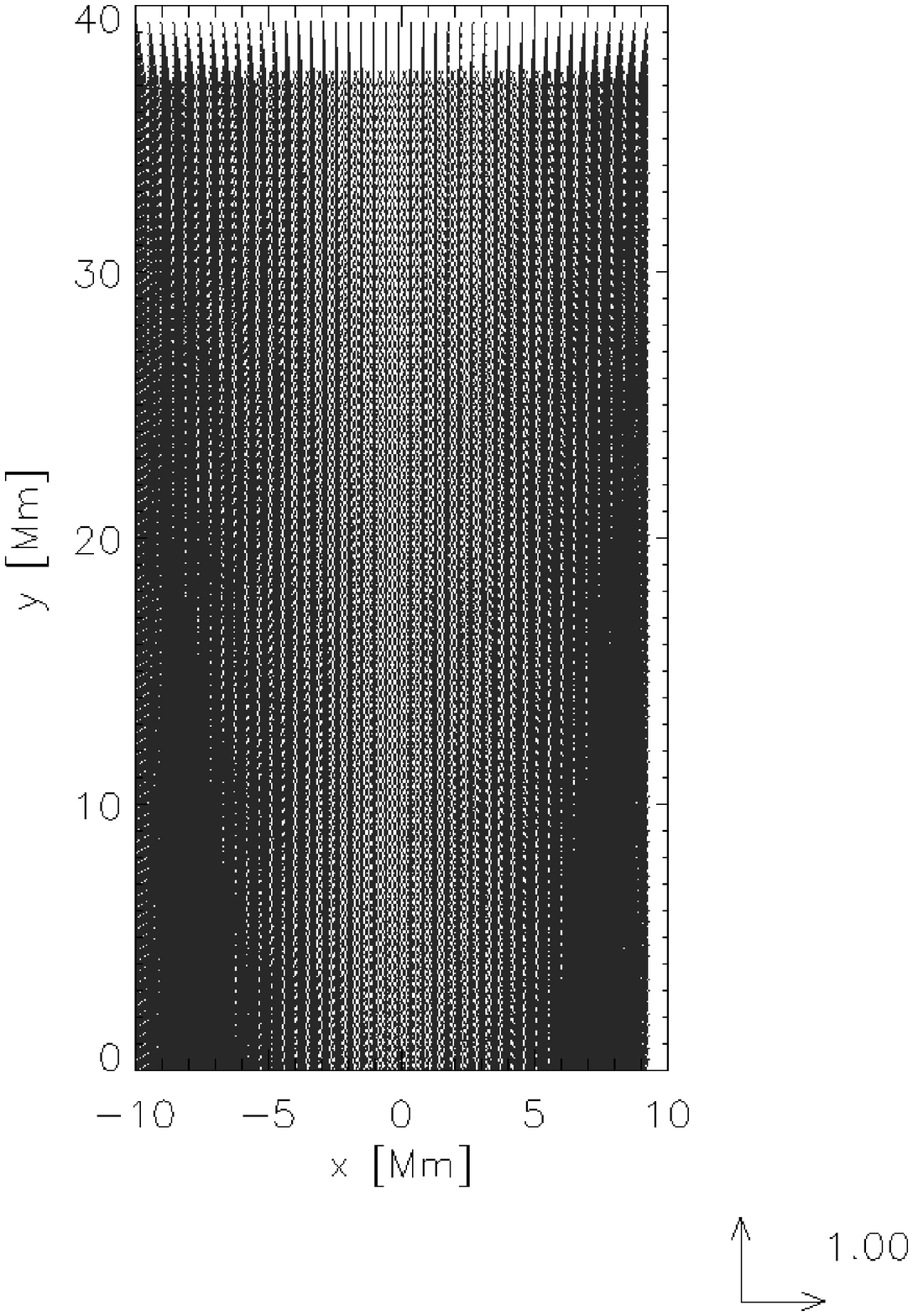}
}
\mbox{
\hspace{-3.2cm}
\includegraphics[scale=0.43]{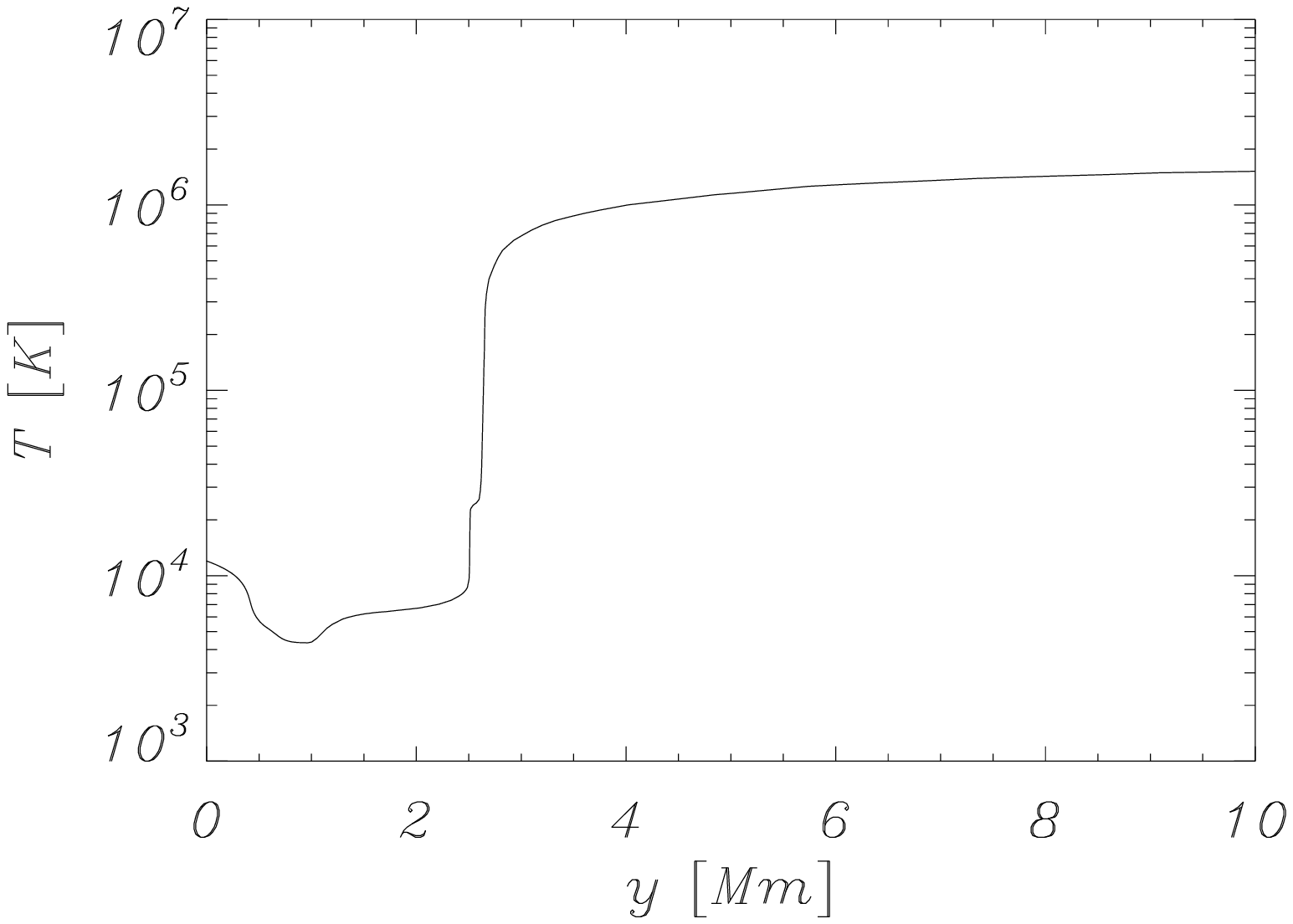}
\hspace{-1.0cm}
\includegraphics[scale=0.43]{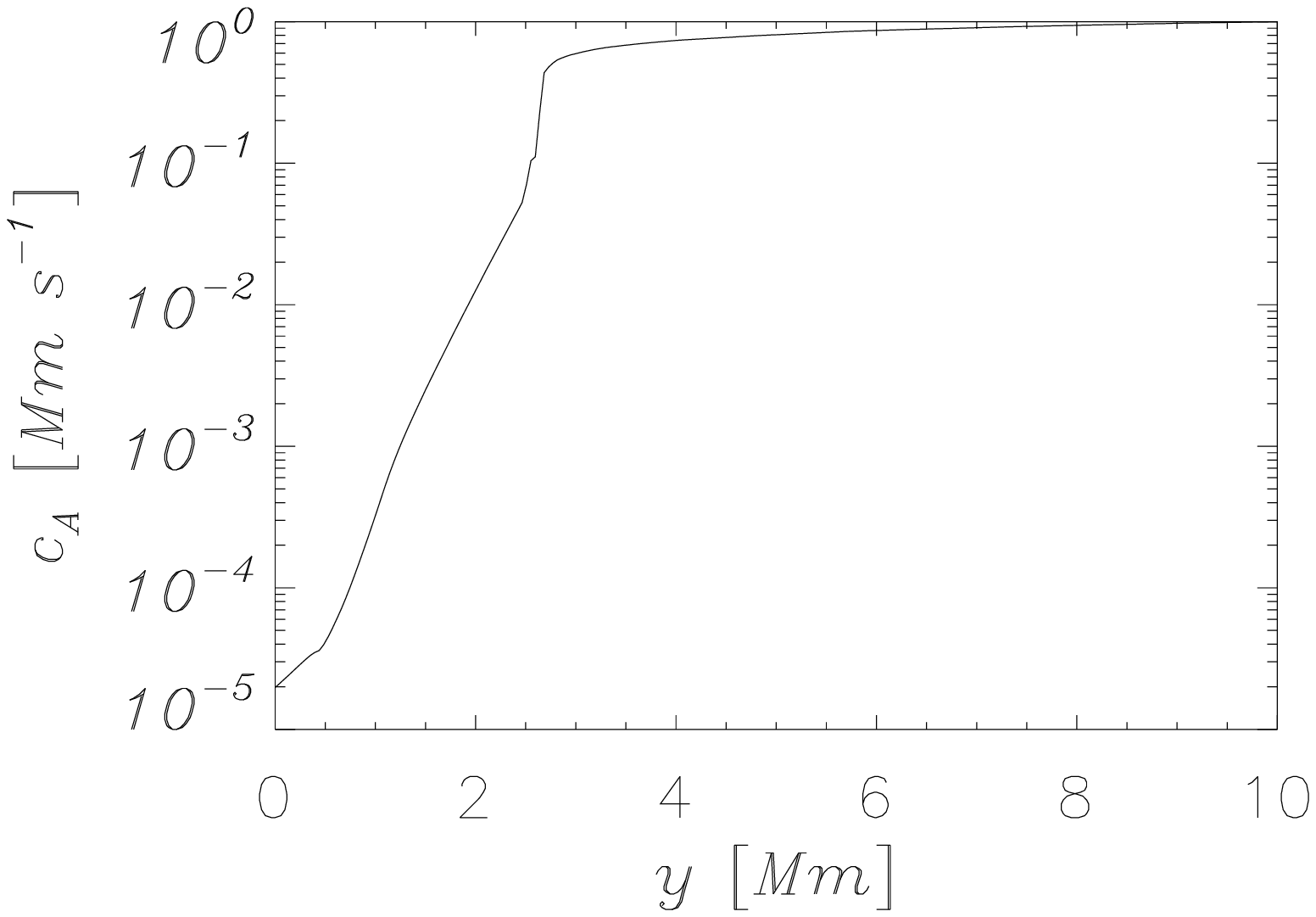}
\hspace{-0.8cm}
\includegraphics[width=0.40\textwidth, height=0.25\textheight]{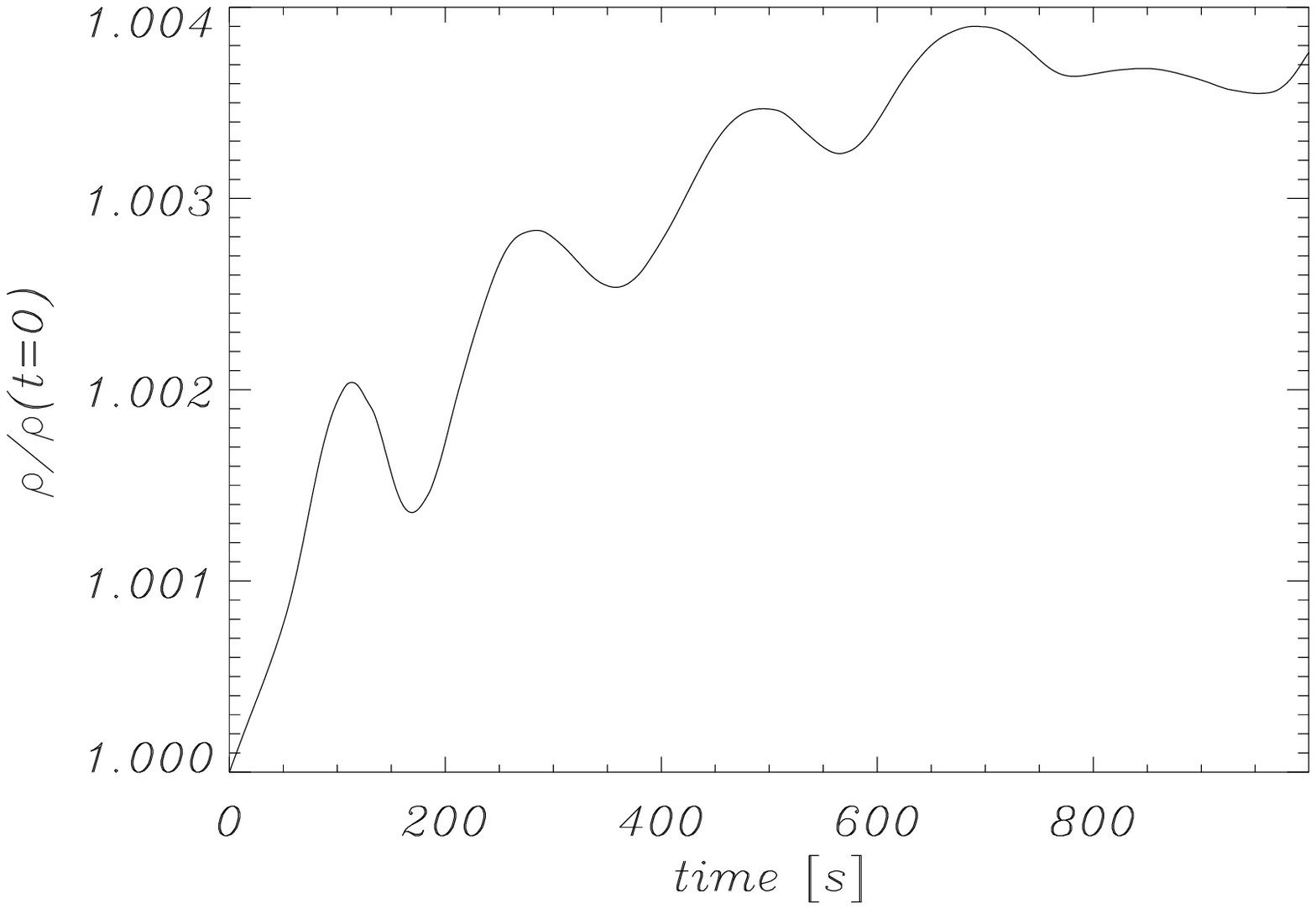}
}
\end{center}
\caption{Top-left panel shows the grid-scheme and blocks of the simulation set-up, while top-right panel 
displays the pattern of the magnetic field vectors.
Bottom-left and bottom-middle panels respectively demonstrate the temperature and Alfv\'en speed 
profiles used in the numerical simulation.Bottom right most panel shows the variation of the normalized mass density collected at a height of 5 Mm in the transition region showing the arrival
of shock wave trains.
}
\label{fig:2}
\end{figure*}
\clearpage
\begin{figure*}
\includegraphics[scale=0.75]{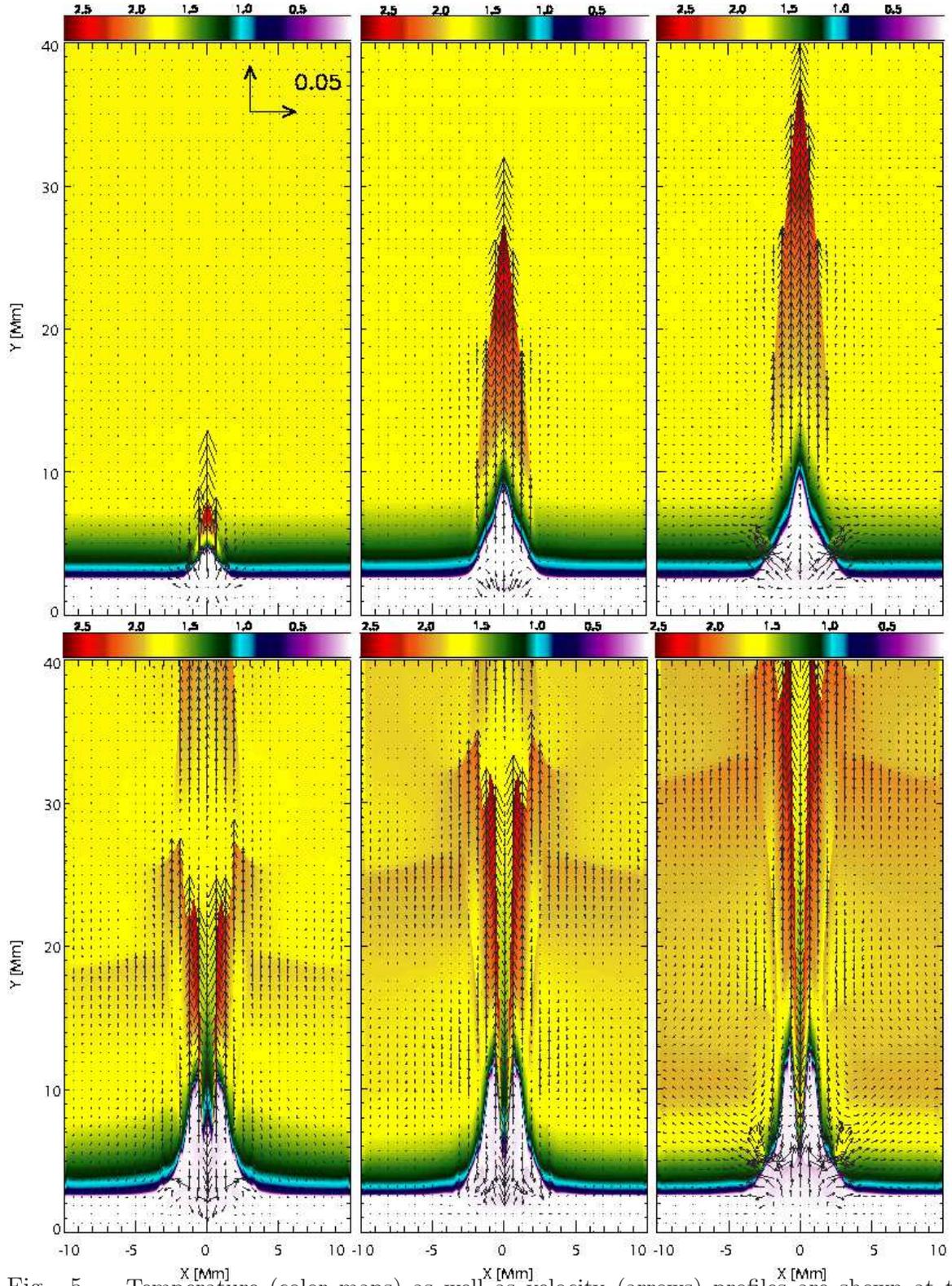}
\vspace{-0.9cm}
\caption{\small Temperature (color maps) as well as velocity (arrows) profiles are shown at t= 100 s, t=200 s, t=250 s, t=400 s, t=450 s, t=500. Temperature drawn in the MK while the arrows represent the velocity expressed in the units of 50 km s $^{-1}$.
}
\label{fig:3}
\end{figure*}
\clearpage

\clearpage

\end{document}